\documentclass{article}

\usepackage{arxiv}

\usepackage[utf8]{inputenc} 
\usepackage[T1]{fontenc}    
\usepackage{hyperref}       
\usepackage{url}            
\usepackage{booktabs}       
\usepackage{amsfonts}       
\usepackage{nicefrac}       
\usepackage{microtype}      
\usepackage{lipsum}
\usepackage{amsmath}
\usepackage{amssymb}
\usepackage{graphicx}

\title{A data-driven approach to precipitation parameterizations using convolutional encoder-decoder neural networks}

\author{
  Pablo Rozas Larraondo\\
  Fenner School of Environment \& Society\\
  ANU Building 141, Linnaeus Way\\
  Canberra, Australia\\
  \texttt{pablo.larraondo@anu.edu.au} \\
  \And
 Luigi J. Renzullo \\
  Fenner School of Environment \& Society\\
  ANU Building 141, Linnaeus Way\\
  Canberra, Australia\\
  \texttt{luigi.renzullo@anu.edu.au} \\
  \AND
  I\~naki Inza \\
  Intelligent Systems Group \\
  University of the Basque Country \\
  Paseo de Manuel Lardizabal, Donostia, Spain \\
  \texttt{inaki.inza@ehu.eus} \\
  \AND
  Jose A. Lozano \\
  Intelligent Systems Group \\
  University of the Basque Country \\
  Paseo de Manuel Lardizabal, Donostia, Spain \\
  \texttt{ja.lozano@ehu.eus} \\
}

\begin{document}
\maketitle

\begin{abstract}
Numerical Weather Prediction (NWP) models represent sub-grid processes using parameterizations, which are often complex and a major source of uncertainty in weather forecasting. In this work, we devise a simple machine learning (ML) methodology to learn parameterizations from basic NWP fields. Specifically, we demonstrate how encoder-decoder Convolutional Neural Networks (CNN) can be used to derive total precipitation using geopotential height as the only input. Several popular neural network architectures, from the field of image processing, are considered and a comparison with baseline ML methodologies is provided. We use NWP reanalysis data to train different ML models showing how encoder-decoder CNNs are able to interpret the spatial information contained in the geopotential field to infer total precipitation with a high degree of accuracy. We also provide a method to identify the levels of the geopotential height that have a higher influence on precipitation through a variable selection process. As far as we know, this paper covers the first attempt to model NWP parameterizations using CNN methodologies.
\end{abstract}


\section{Introduction}

Numerical Weather Prediction (NWP) currently underpin most weather forecasting operations across the globe. The steady increase in NWP skill over the past 40 years is linked to the growth in computing power, the increased availability of observational data, and the development of better data assimilation methods \cite{bauer2015quiet}. However, despite technological advances, some important physical processes, such as convection, friction, turbulence or radiation, manage to elude adequately representation at the sub-grid scales for most NWP systems \cite{kalnay2003historical,stensrud2009parameterization}. NWP uses parameterizations to model these physical processes, which rely on the availability of explicitly resolved parameters. Parameterizations are often based on empirical assumptions used to represent processes which occur at sub-grid scales and constitute a major source of uncertainty in NWP \cite{palmer2005representing,slingo2011uncertainty}.

Most parameterizations are based on deterministic model equations representing simplifications of the physical processes. There are also some that are based on statistical or probabilistic approaches. These parameterized processes can interact with each other leading to complex model behaviours. Often, a small modification in one of its components can lead to inconsistencies with other parameterizations and ultimately results in instabilities in NWP estimation. The process of designing and maintaining individual parameterizations and the relationships between them, is therefore laborious and requires a high level of domain-specific knowledge.

The ready access to large volumes of data permits the use of data-driven methods to derive parameterizations\cite{berner2017stochastic}. Machine Learning (ML) has been proposed to derive parameterizations using methods such as regression trees \cite{belochitski2011tree,o2018using} or neural networks \cite{krasnopolsky2013using,brenowitz2018prognostic}. In meteorology, the question of whether deep neural network models \cite{lecun2015deep}, for example, trained on atmospheric data, can compete with physics-based NWP models has been recently addressed demonstrating promising results \cite{dueben2018challenges,scher2018towards}. Similar deep learning networks have also been explored in the context of parameterizing sub-grid physical processes in NWP. For example, deep neural networks have been used to perform prognostic simulations \cite{brenowitz2018prognostic} or to learn convective and radiative processes from cloud resolving models \cite{rasp2018deep} using single column models.

In this paper, we propose the use of a specific kind of deep neural network, called an encoder-decoder convolutional neural network (CNN), to learn relationships between NWP variables and demonstrate their potential as a novel approach to parameterization. Specifically, we demonstrate how the geopotential height parameter can be used exclusively to infer total precipitation -- a field that is strongly dependent on NWP parameterizations and other NWP explicitly resolved parameters such as humidity or temperature. Our objective here is to demonstrate that despite only using the most basic of NWP fields, i.e. geopotential height, CNNs have the capacity to extract the necessary information to estimate precipitation in the absence of additional physical fields. NWP offers fields closely related to precipitation, such as humidity or temperature. However, the objective of this work is to demonstrate the capacity of CNNs to extract the information contained in geopotential height, which is one of the most basic fields.

The challenge is to design a model that is capable of extracting the synoptic and mesoscale spatial information contained in the geopotential height and that finds the relationships with the corresponding total precipitation field. To train the networks for the experiments presented in this manuscript, we use 40 years of reanalysis data made available by ECMWF's ERA-Interim model \cite{dee2011era}.  

We demonstrate that encoder-decoder CNNs are able to learn, with some skill, the relationship between geopotential height and precipitation. We devise a methodology to identify the input variables (atmospheric levels in our case) which result in the most accurate total precipitation estimates. We also compare the skill of several encoder-decoder architectures, from the domain of computer vision, identifying the most accurate architecture for deriving precipitation.

The paper is structured as follows: Section 2 provides an overview of CNNs, identifying the key components relevant to NWP learning and describing the specific class of CNN used in our research. Section 3 presents the dataset and methodology used in the experiments. Section 4 shows the results of the experiments offering a comparison with traditional methodologies, including linear regression and random forest for a selected group of locations across the extended European region. We finish with Section 5, which provides conclusions and ideas on how the proposed methodology can be further developed in the future.

\section{Convolutional neural networks in the context of weather forecasting}

NWP outputs are expressed as two-dimensional (2d) numerical arrays of numbers for any given time at discrete levels of the atmosphere. Typically these 2d weather fields are represented as digital images comprising the latitude and longitude dimensions. In this section we introduce Convolutional Neural Networks (CNN), a popular image analysis methodology from the field of computer vision. We describe their application to NWP and parameterization problems, referring to 2d arrays of numerical data as \emph{images} hereafter.

\subsection{Image convolution}
Convolution, in the context of image processing, is a mathematical operation performed over a  neighbourhood of pixels in an image, weighted by a 2d matrix called a \emph{kernel}. The output of convolution operation is another transformed image with the same dimensions as the original image (ignoring border effects). Performing image convolution requires the application of the same operation iteratively, by sliding the kernel across the whole image. The result of each convolution operation is assigned to the pixel in the new image at the position designated by the kernel's centre.
The following equation shows the decomposed convolution operation between the different grid points in a region of a weather field \textbf{F}, a convolution kernel \textbf{K} and how the results are assigned to an output image \textbf{C}: 

\begin{equation}
c_{ij} = 
\begin{bmatrix}
    f_{i-1,j-1} & f_{i-1,j} & f_{i-1,j+1} \\
    f_{i,j-1} & f_{i,j} & f_{i,j+1}\\
    f_{i+1,j-1} & f_{i+1,j} & f_{i+1,j+1}
\end{bmatrix}
\circledast
\begin{bmatrix}
    k_{11} & k_{12} & k_{13}\\
    k_{21} & k_{22} & k_{23}\\
    k_{31} & k_{32} & k_{33}
\end{bmatrix}
\end{equation}

\begin{figure*}[h]
 \centerline{\includegraphics[width=10cm]{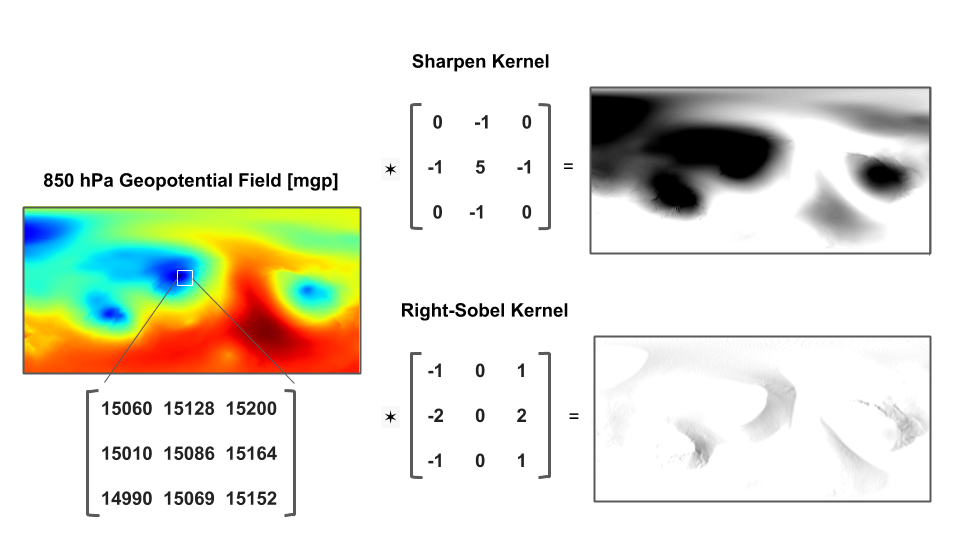}}
  \caption{Illustration of the convolution operation applied to image of 850 hPa geopential heights. Two kernels are used: a sharpen kernel and a edge-detecting kernel. The resulting output of the operation is displayed using greyscale images.}\label{convolution_op}
\end{figure*}

Figure \ref{convolution_op} contains a representation of how a convolution operation is applied using a $3 \times 3$ kernel over a region of a gridded NWP 850 hPa geopotential field. Two kernels are used: a smoothing kernel (top) and a edge-detecting kernel (bottom). 
The figure shows the effect of two different convolution kernels on a 850 hPa geopotential height field as grey-scale images. The Right-Sobel kernel, for example, detects the edges with an orientation to the right or, in other words, it makes salient the regions of the image that contain right to left decreasing gradients. We observe that this kernel has identified features in the geopotential that are reminiscent of those that a trained weather forecaster might identify as frontal systems. This automated method for the identification of key features provides a targeted approach for identifying weather phenomena over vast areas and large volumes of numerical data.

\subsection{Convolutional neural networks}
Convolutional Neural Networks (CNN's) \cite{lecun2010convolutional} contain a series of layers which perform convolution operations, extracting the spatial information of input images. Instead of using predefined kernels, CNN's use gradient descent method \cite{bottou2010large} to find the optimal matrix weight values that reduce a loss function. This loss function measures the accuracy of a task during training. The network learns, during the training process, the weights that transform a set of input images into a close representation of its corresponding outputs.

\begin{figure*}[h]
 \centerline{\includegraphics[width=10cm]{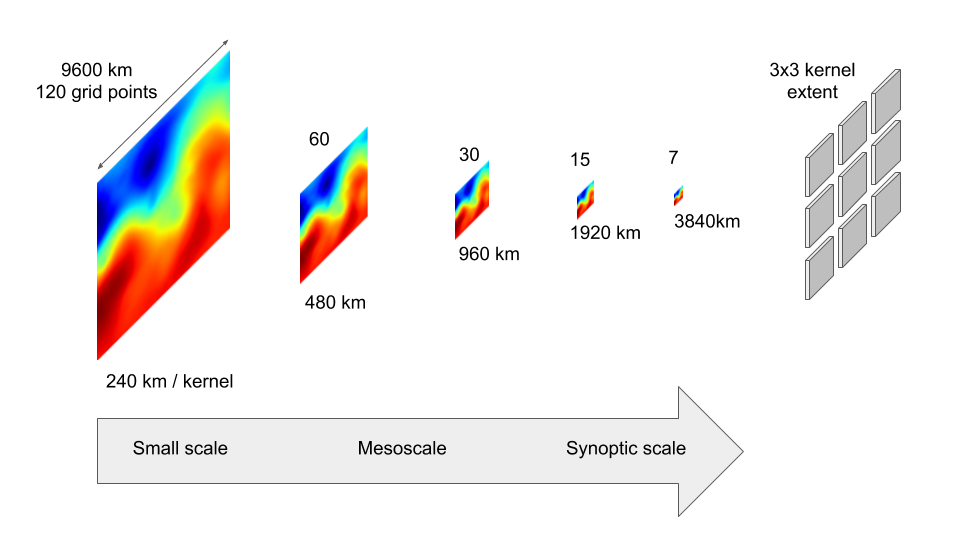}}
  \caption{Graphical representation of the dimensionality reduction performed by each convolutional layer in the network. A fix sized kernel captures a progressively larger area in each layer.}\label{cnn_scales}
\end{figure*}

Convolutions, at each layer of a CNN, are commonly followed by an operation that causes a reduction in the size of the image, and an activation function to introduce non-linearities in the model \cite{glorot2010understanding}. The dimensionality reduction is usually achieved by applying a specific subsampling method called pooling \cite{scherer2010evaluation} or by performing the convolution operation on strides over the image. The convolution, pooling and non-linear activation function, performed at each layer of a CNN, result in a gradual reduction in the size of the input image, which implies that kernels can cover increasingly larger areas relative to the initial image. This characteristic allows CNNs to identify features at different scales in an image. Fig \ref{cnn_scales} represents the reduction in the size of a NWP geopotential height field performed by a 4 layers CNN. Each pooling operation in this network reduces the size of the image to a half. As the dimensions of the image decrease, each grid point represents a larger area and kernels are able to extract mesoscale and synoptic scale features in the initial image. 

Each layer in a CNN transforms the dimensionality of the input image by reducing its spatial extent and expanding the number of features along a third dimensional axis. This expansion in depth of the images is achieved using 3-dimensional kernels, by stacking 2-dimensional kernels, each detecting specific patterns or features of the input image.

CNNs are an extremely popular and effective method to solve classification and regression tasks. Although these networks were originally developed in the field of computer vision, they have found applications in other fields, such as medical image analysis \cite{litjens2017survey}, high-energy physics \cite{baldi2014searching} or astronomy \cite{dieleman2015rotation}. 

\subsection{Convolutional encoder-decoder networks}

Autoencoders \cite{hinton2006reducing} are a type of neural networks that can learn to create compressed representations of the input data. These networks perform a dimensionality reduction followed by an expansion of the input space trained to recreate the original input. Convolutional autoencoders \cite{masci2011stacked} use CNNs to learn compressed representations of images. Convolutional encoder-decoder networks use convolutional layers to compress, or reduce the dimensionality of input images, followed by deconvolution layers, which perform the inverse operation by expanding the dimension of images. The encoder and decoder parts define a symmetrical structure in which images get compressed into a latent representation and then decompressed into its original dimensions. Convolutional encoder-decoder networks have been mainly applied to image segmentation tasks, in which the network identifies the pixels of an image belonging to the same class \cite{krizhevsky2012imagenet,long2015fully,chen2018deeplab}.

A generalisation of convolutional encoder-decoder networks is found in pixel-to-pixel regression or image-to-image translation networks \cite{isola2017image}, where similar architectures are used to perform regression between images instead of classification. The same networks used to perform image segmentation tasks can be applied to regression problems by modifying the network's loss function. Using a regression loss function, such as Root Mean Square Error (RMSE) or Mean Absolute Error (MAE), these networks can learn to predict pixel values in a continuous domain.

\begin{figure*}[h]
 \centerline{\includegraphics[width=13cm]{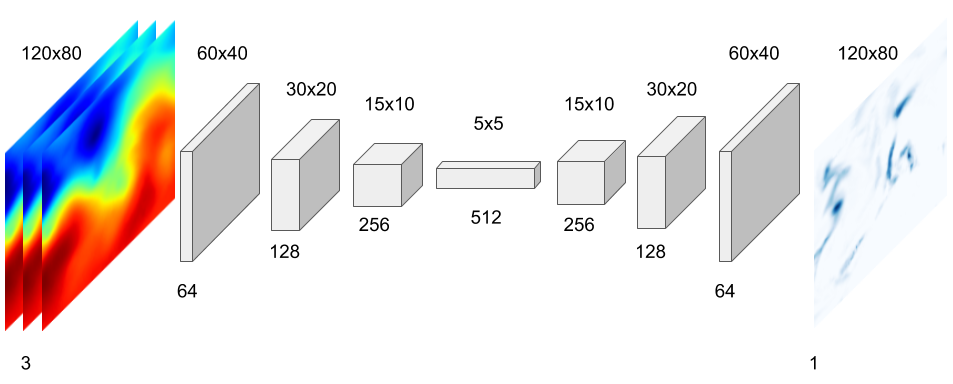}}
  \caption{Transformations in the dimensionality of the data performed by a VGG-16 encoder-decoder to map between the input and output spaces.}\label{vgg16}
\end{figure*}

In the context of weather forecasting, convolutional encoder-decoder networks can be used to learn the relationship between physical variables. These networks are modified, in this paper, to perform regression and find the relationship between the NWP geopotential height and total precipitation. Figure~\ref{vgg16} represents the basic structure of the VGG-16 convolutional encoder-decoder network showing the transformations performed to the dimensionality of the data. In this example, the network creates a mapping between two spaces defined by the input geopotential height and the output total precipitation. 

The main constraint of convolutional encoder-decoder networks is the loss of spatial information, caused by the dimensionality reduction pooling operations \cite{scherer2010evaluation} on the encoder half. This information cannot be fully recovered by the decoder section and as a consequence, the output images often become blurry or contain ill defined features.

For this work we consider three different state-of-the-art convolutional encoder-decoder networks in the field of image segmentation: VGG-16 \cite{long2015fully}, Segnet \cite{badrinarayanan2017segnet} and U-net \cite{ronneberger2015u}. These three convolutional encoder-decoder networks perform similar dimensionality transformations to the data. The difference between them resides in the number of convolution operations performed at each layer and in the configuration of the connections between layers. VGG-16 presents a linear architecture, in which each layer is only connected to the adjacent ones whereas the other two networks present features to improve the quality of the reconstructed images. Segnet computes the index of the max pooling operation at each encoder layer and communicates this value to its symmetric in the decoder, so they can be used in the up-sampling stage. The U-net decoder, on the other hand, concatenates the weights used in the encoder part to reconstruct the spatial information at the decoding stage. These last two network address therefore the problem of preserving the spatial information lost during the encoding stage.

NWPs produce a large number of inter-related physical variables with different levels of dependence. Some of these variables are derived from other fields using physical equations and others are parameterized. In this context, convolutional encoder-decoder networks provide a generic methodology to perform regression between different weather fields, learning the underlying relationships between them. In the next sections we explore the use of convolutional encoder-decoder networks to learn NWP parameterizations.

\section{Dataset and methodology}

\subsection{Dataset}
\begin{figure*}[h]
 \centerline{\includegraphics[width=13cm]{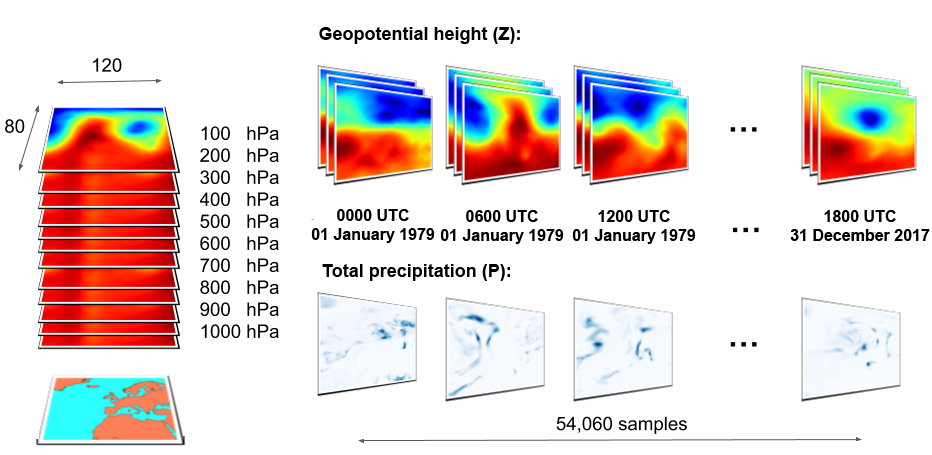}}
  \caption{Representation of the geographic study area (\textit{latitude: [75, 15], longitude = [-50, 40]}) and temporal extent covered by ERA-Interim geopotential height and total precipitation.}\label{dataset}
\end{figure*}

Here we use the NWP ERA-Interim \cite{dee2011era} global climate reanalysis dataset produced by the European Centre for Medium-Range Weather Forecasts (ECMWF). ERA-Interim contains reanalysis data from 1979 to present with a 6-hour temporal resolution. The spatial resolution of the dataset is approximately 80 km (reduced Gaussian grid N128) on 60 vertical levels from the surface up to 0.1 hPa pressure level. ERA-Interim data is publicly accessible from ECMWF's Public Datasets web interface \cite{1321008426928}. From the large volume of output variables available, we choose geopotential height \textit{(\textbf{Z})} and total precipitation \textit{(\textbf{P})}. In addition, we focus on the mid-latitudes region defining a rectangular area over Europe bounded by (\textit{latitude: [75, 15], longitude = [-50, 40]}). Figure \ref{dataset} represents the geographic area as well as the correspondence between the geopotential height and the total precipitation field time series.

Note that we select a subset of the available geopotential heights \textit{(\textbf{Z})} corresponding to the following pressure levels of the atmosphere: \{1000, 900, 800, 700, 600, 500, 400, 300, 200, 100\} hPa. The resulting geopotential height data are stored as a 4-dimensional numerical array with shape [54.023, 10, 80, 120] for the corresponding dimensions [time, height, latitude, longitude].

The ERA-Interim \textit{(\textbf{tp})} represents an accumulated amount over a 3-hour period for each grid point. This field is further aggregated to match the 6-hour frequency of the geopotential height field. The result is a 3-dimensional numerical array with shape [54.023, 80, 120] for the corresponding dimensions [time, latitude, longitude].

\subsection{Experimental design}

The objective of the experiments described here is to demonstrate that convolutional encoder-decoder neural networks can learn to infer complex atmospheric processes, such as precipitation, using basic NWP fields as input. We consider 3 state-of-the-art convolutional encoder-decoder neural networks to find the relationships between geopotential height and total precipitation comparing their performance.

To facilitate the comparison between the different encoder-decoder models, we propose the use of a pipeline comprising two steps. The first step performs a variable selection process to determine the geopotential height levels that minimise the error at forecasting the total precipitation field. The second step compares the results of three state-of-the-art convolutional encoder-decoder architectures, in the field of image segmentation, at learning to diagnose the ERA-Interim total precipitation field.

\subsubsection{Variable selection}

The input dataset comprises ten levels of the geopotential height; however, due to the linear increase in the number of trainable parameters and the hardware requirements to train these convolutional encoder-decoder networks, we limit the number of input levels to three. Different methods for performing variable selection have been proposed \cite{saeys2007review}. These methods generally reduce the search space of input variables and optimize the construction of accurate predictors. 

For our experiment, we build a simplified convolutional encoder-decoder network and perform an exhaustive search over all the possible combinations of the considered geopotential height levels. This simplified network has a similar architecture to the deeper encoder-decoder networks, but its complexity is reduced by limiting the number of layers and depth of the convolution operations. Performing a similar exhaustive search of input variables with the deeper, more complex networks, like those introduced in the next part of the experimental process, would be computationally infeasible. However, this simplified network allows a quick iteration across the whole feature space to identify the subset of geopotential heights that minimizes the error of the precipitation field in the training set.

To identify the levels of geopotential height that produce the best precipitation results, the simplified convolutional encoder-decoder network is trained over the 1-, 2- and 3-combinations out of a set of 10 pressure levels, which results in 175 combinations.  
The results of this variable selection process are used to compare the accuracy of the three state-of-the-art deep encoder-decoder convolutional networks in the second step of the pipeline. Therefore, the final validation comparing the accuracy of the different networks must be performed using a different dataset partition that remains unseen during the variable selection process  \cite{reunanen2003overfitting}.

\begin{figure*}[h]
 \centerline{\includegraphics[width=13cm]{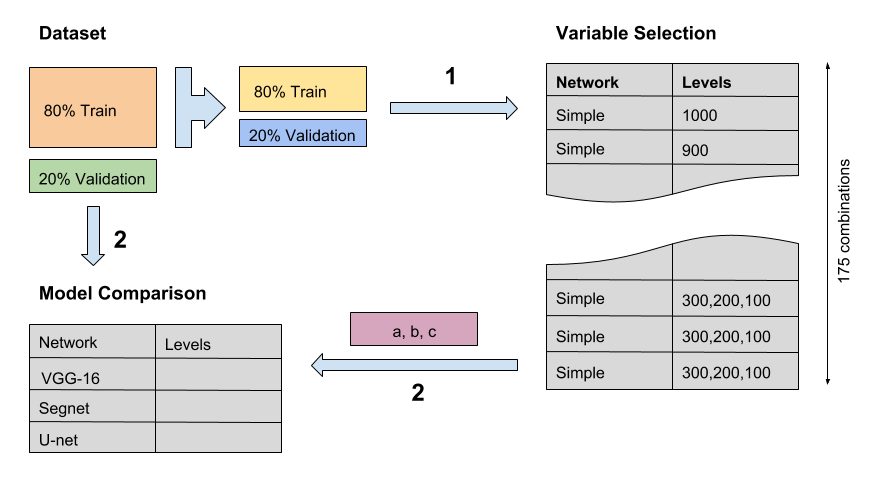}}
  \caption{Graphical representation of the experiment pipeline comprising the variable selection and the encoder-decoder network comparison processes.}\label{exp}
\end{figure*}

The whole pipeline can be seen as a variable selection process followed by the training process of the three compared networks. The initial dataset, which contains 54,060 samples, is randomly split into the training and validation datasets, containing 80\% and 20\% of the data respectively, so that the different meteorological situations are evenly represented in both splits. These partitions are used to evaluate the differences in accuracy between the compared deep architectures. The variable selection process is performed internally using the training dataset, which is further split into 80\% and 20\% internal partitions to train and validate the different subsets of input variables. The final comparison between the architectures is performed using the initial outer 20\% validation split, which does not intervene at neither the variable selection process nor the training of the different architectures. Figure \ref{exp} represents the proposed experiment and the relationship between the variable selection and the model evaluation processes. 

The network used in the variable selection process is a simplification of the VGG-16 convolutional encoder-decoder network \cite{long2015fully}, which has demonstrated state-of-the-art results in image segmentation tasks. Figure~\ref{vgg16} represents the dimensionality transformations performed by the VGG-16 network to the input space using 3x3 convolutions using stride value of 2 to perform the spatial dimension reduction. The simplification proposed for this part of the experimental process consists in removing the last convolution operation in the encoder section. The information gets compressed to a depth of 256 channels instead of the 512 in the full VGG-16 network. This simplified network reduces significantly the total number of parameters when compared to the full VGG-16 which results in a significant reduction in the amount of compute resources required to train it. Each network is trained during 20 epochs (iterations over the internal input training split) and their results are computed using the MAE metric comparing the output to the total precipitation field in the internal validation dataset. The results of this first experiment are therefore used to determine the geopotential height levels that minimize the error at forecasting total precipitation.

\subsubsection{CNN model selection}
The second step of the pipeline focuses on training each deep encoder-decoder convolutional network to forecast ERA-Interim's total precipitation field. The objective of this second half of the experimental process is to determine the network that provides the best accuracy when forecasting total precipitation. For this part, we use the levels of geopotential height, selected previously, to compare the three different state-of-the-art segmentation networks.

We consider three different state-of-the-art convolutional encoder-decoder networks in the field of image segmentation: VGG-16 \cite{long2015fully}, Segnet \cite{badrinarayanan2017segnet} and U-net \cite{ronneberger2015u}. These networks are modified to perform regression tasks instead of classification by changing the loss function to MAE. To accomplish an honest comparison between these three networks, we build them using the same number of layers and depth of the convolution operations. The basic structure for all three networks is represented in Figure~\ref{vgg16}. Therefore, all three networks perform the same dimensionality transformations when estimating total precipitation from the geopotential height input. The difference between these networks resides in the number of convolution operations performed at each layer and in the configuration of the connections between layers. VGG-16 presents a linear architecture, in which each layer is only connected to the adjacent ones. Segnet computes the index of the max pooling operation at each of the encoder layers and communicates this value to its symmetric in the decoder, so they can be used in the up-sampling stage. The U-net decoder, on the other hand, concatenates the weights used in the encoder part to reconstruct the spatial information.

All three networks are trained using the initial outer 80/20 split defined at the beginning of the experiment. This way, the validation split used to compare the accuracy of the different networks has remained unseen during the variable selection and training of the different networks. This method assures the independence and fairness of the results between the results from each part of the pipeline. 

The networks are trained during 50 epochs using the subset of geopotential heights selected in the first part of the experiment as input and the total precipitation field as output. The same optimiser (stochastic gradient descent,\cite{bottou2010large}), learning rate (0.01) and loss function (MAE) as in the variable selection process are used to train the three networks. These networks are then compared with the total precipitation field in the validation split dataset produced by ERA-Interim, to determine the error.

The models are implemented using the Keras \cite{chollet2017keras} framework and the TensorFlow \cite{abadi2016tensorflow} back-end. These models, as well as a copy of the dataset used in the experiments, are available at this repository: \url{https://github.com/prl900/precip-encoder-decoders}.

\section{Results and discussion}

\subsection{Variable selection process}

In the first part of the experimental process we identify a subset of the geopotential height levels that produce a better estimate in the training set of the ERA-Interim total precipitation field. We train the simple encoder-decoder network 175 times as described in the previous section using the training split. 

\begin{figure}[h]
 \centerline{\includegraphics[width=8cm]{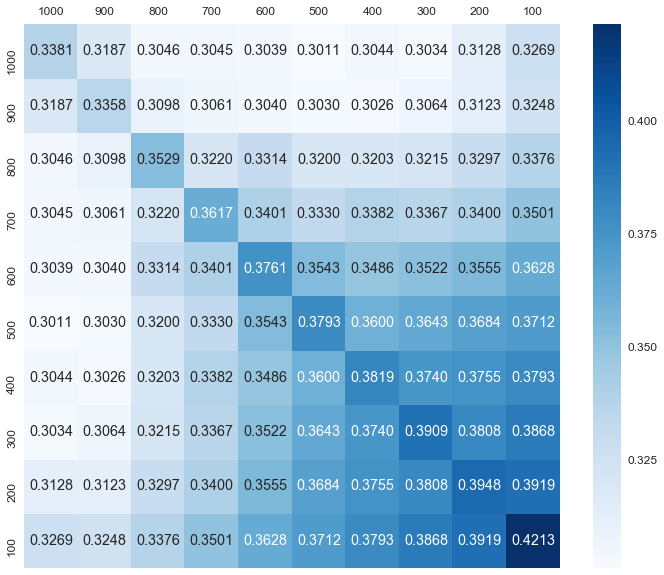}}
  \caption{Matrix representing the average validation MAE results for each simple encoder-decoder network trained with each possible combination of two geopotential levels.}\label{heatmap}
\end{figure}

The resulting models are then compared using the internal validation split, which is formed with the remaining 20\% of the initial training split. The MAE metric is used to compare the error in diagnosing total precipitation at each point of the grid. Figure \ref{heatmap} contains the MAE scores produced for each combination of any two geopotential height levels relative to the total precipitation field produced by the NWP. The results indicate that combinations of lower levels of the atmosphere produce better estimates of the precipitation field than the higher levels. The main diagonal of the matrix in Figure \ref{heatmap} represents the resulting errors when using a single geopotential level to train the network. Individually, the lower levels of the atmosphere present lower MAE values when forecasting precipitation, being 900 hPa the one with the lowest error. In the case of using two inputs, the lowest errors are found when combining low and mid levels of geopotential, being the combination of 1000 and 500 hPa levels the one that presents the lowest error.

Training the encoder-decoder network with three geopotential levels, results in a significant improvement in performance. Table \ref{3leveltable} contains the five lowest error results and their corresponding atmospheric levels. Unfortunately, the results for three levels cannot be easily represented graphically. Compared to the previous results, there is a notable improvement in performance when a third level is added as input to the encoder-decoder network. This implies that the neural network is capable of finding internal relationships between the different levels of the atmosphere and relate them with precipitation events. The combination of 1000, 800 and 400 hPa geopotential heights results in the lowest error of the total precipitation field in the training partition. This result is surprisingly similar to the traditional practice in weather forecasting of using 850 and 500 hPa geopotential fields (in conjunction with others, such as temperature and wind) to determine the location of weather fronts and therefore, precipitation \cite{sanders1995case,sanders1999proposed}.

\begin{table}[h]
\caption{Top 5 average MAE results when training the simple encoder-decoder network with every combination of three geopotential height levels to diagnose the ERA-Interim total precipitation field.}\label{3leveltable}
\begin{center}
\begin{tabular}{cc}
$z\hspace{1em}levels$ [hPa]& $MAE$ [mm]\\
 1000, 800, 400 & 0.2895 \\
 1000, 800, 500 & 0.2897 \\
 1000, 900, 500 & 0.2897 \\
 1000, 900, 400 & 0.2901 \\
 1000, 700, 400 & 0.2927 \\
\end{tabular}
\end{center}
\end{table}

The mean absolute error (MAE) values represented in Table \ref{3leveltable} are calculated using the average of the MAE results over the $120 \times 80$ grid area and for all the temporal entries in the validation partition. Considering that the total precipitation field is expressed in millimetres of liquid-equivalent precipitation, the error of these networks when forecasting total precipitation is, on average, less than 1/3th of litre per square metre in a 6-hour period, when compared to the values produced by the NWP.

\subsection{Comparing deep convolutional networks}

For this part of the experimental process, we choose the subset of 1000, 800 and 400 hPa geopotential heights to evaluate the performance of the previously introduced deeper, state-of-the-art segmentation encoder-decoder networks adapted to perform regression tasks. The number of parameters and depth of these networks is significantly higher than the simplified network previously used to perform the selection of the geopotential levels. Training these deeper networks demands therefore significantly higher compute and memory resources. We use a compute node equipped with a NVIDIA Tesla P100-PCIE-16GB Graphical Processing Unit (GPU) provided by the Australian National Computational Infrastructure.

Table \ref{perftable} represents the total number of trainable parameters for each of these networks and the total amount of time required to train the different networks during 50 iterations (epochs) over the initial training split. The total number of trainable parameters provides an indication of the size of each network and the time value in this table gives an indication of the time required to train each network using TensorFlow \cite{tensorflow2015-whitepaper}, an open-source library released by Google, and P-100 GPU nodes.

\begin{table}[h]
\caption{Number of parameters for each encoder-decoder network architectures and resulting training time for each network (50 epochs).}\label{perftable}
\begin{center}
\begin{tabular}{crc}
$Network$ & $Parameters$ & $Time\ [hours]$\\
 Simple (ref.) & 745,000 & 0.6 \\
 VGG-16 & 16,467,469 & 4.7 \\
 Segnet & 29,458,957 & 8.6 \\
 U-net & 7,858,445 & 2.4 \\
\end{tabular}
\end{center}
\end{table}

\begin{figure}[h]
 \centerline{\includegraphics[width=8cm]{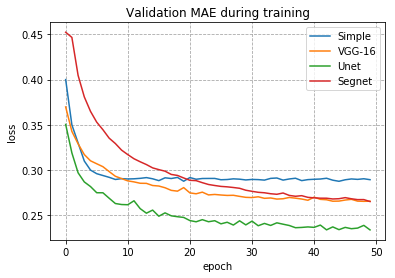}}
  \caption{Comparison of the evolution of the validation error during training for the four convolutional encoder-decoder networks over 50 epochs.}\label{training}
\end{figure}

Figure \ref{training} shows the learning process of the four different encoder-decoder networks during 50 epochs over the training dataset. At the end of each epoch during training, the validation dataset is used to assess the error of the model and the improvement of the different models can be compared using unseen data. At the beginning of the training process the network learns fast and it slows down as the training progresses. The reduction in the validation error is different for each network which flattens at different points and rates.

\begin{figure*}[h]
 \centerline{\includegraphics[width=17cm]{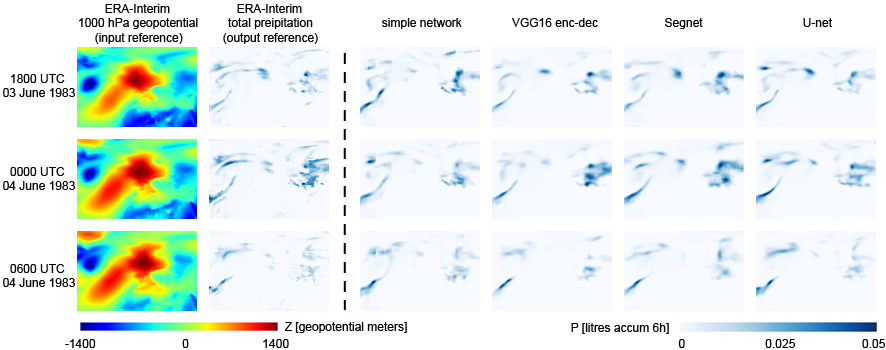}}
  \caption{Visual comparison between the total precipitation field generated by the different networks. ERA-Interim 1000 hPa geopotential height and total precipitation fields are included for reference.}\label{comparison}
\end{figure*}

\begin{table}[h]
\caption{Accuracy of the different networks using the validation split at the end of the training process.}\label{deep_results}
\begin{center}
\begin{tabular}{cl}
$Network$ & $MAE$\ [mm]\\
 Simple & 0.2893 \\
 VGG16 & 0.2630 \\
 Segnet & 0.2618 \\
 U-net & 0.2386 \\
\end{tabular}
\end{center}
\end{table}

Considering the validation results in Figure \ref{training} and the size of each network in Table \ref{perftable}, we highlight the behavior of U-net which shows the lowest validation error and is also the one with the lowest number of parameters of the three deep learning networks considered. 

Figure \ref{comparison} offers a visual comparison between the outputs generated by each model for an atypical atmospheric situation around 0000 UTC 04 June 1983. The first two columns from the left represent the 1000 hPa geopotential height and total precipitation, as produced by the ERA-Interim model. Total precipitation represents the total precipitation accumulated over the 6-hour period following the indicated time. In a similar way, and using the same colour scale, the next 4 columns represent the precipitation generated by the different encoder-decoder networks.

The spatial structure and intensity of the precipitation field is represented differently by each network, with slight variations in respect of the ERA-Interim reference output. Different convolutional encoder-decoder networks use different methods to reconstruct the spatial information lost during the encoding phase. Apart from capturing the spatial structure of the precipitation field, the different networks provide accurate results for the precipitation intensity at each grid point.

\subsection{Statistical analysis of the results}
\begin{figure}[h]
 \centerline{\includegraphics[width=8cm]{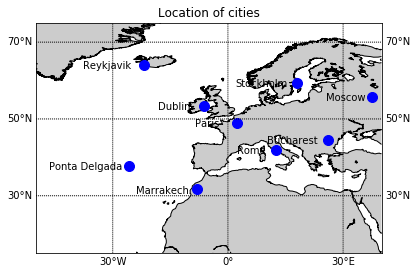}}
  \caption{Location of the nine different cities within the comprised region.}\label{cities}
\end{figure}

To compare the performance of each convolutional encoder-decoder architectures, we use the external validation split to extract the total precipitation at the closest grid point to nine different cities. Figure \ref{cities} represents the geographical location of these nine cities within the region of our dataset. These cities are located in different climatic zones and present distinct precipitation patterns.

\begin{figure*}[h]
 \centerline{\includegraphics[width=16cm]{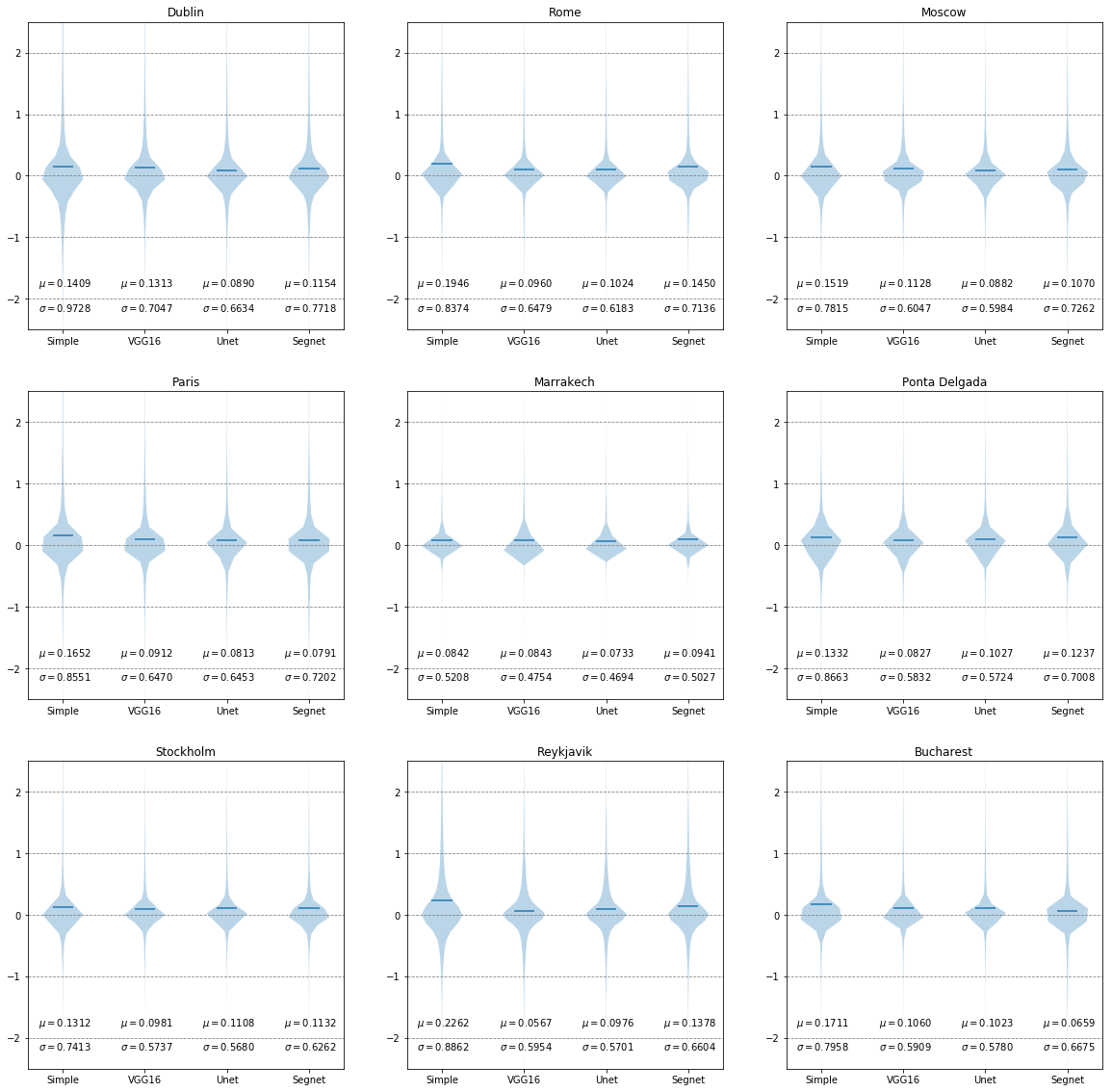}}
  \caption{Representation of the error density function and the mean $(\mu)$ and standard deviation $(\sigma)$ values for the different neural network architectures at each city.}\label{violin}
\end{figure*}

Results are assessed using the ERA-Interim total precipitation field as reference for the same grid points using the signed error -- or bias -- metric. This metric provides information about possible biases and distribution of the error as opposed to the previously used MAE, which does not provide information about the sign of the error. For each city and point in time the error at estimating total precipitation is calculated. These results are then aggregated by city and type of network. Figure \ref{violin} uses a violin plot \cite{hintze1998violin} to represent the error results at each location for the different architectures. A violin plot proposes a modification to box plots adding the density distribution information to the basic summary statistics inherent in box plots. The horizontal blue bar towards the centre of each of the violins in Figure \ref{violin} represents the mean. The lower part in each plot shows the mean $(\mu)$ and standard deviation $(\sigma)$ of the error values for each network and location. The shape of the violin gives a visual indication of each model's performance. Wider and sharper violin shapes around the 0 value provide an indication of good network performance.

In order to statistically compare the results, we use the methodology proposed by \cite{demsar2006} to assess the statistical significance of the differences between the error results of each network across the nine locations. The initial Friedman test rejects the null hypothesis of similarity among the 4 convolutional encoder-decoder networks. This justifies the use of post-hoc bivariate tests, Nemenyi \cite{pohlert2014pairwise} in our case, to assess the significance of the differences between the different pairs of encoder-decoder networks. 

The results of these tests are graphically expressed using Critical Difference (CD) diagrams. The Nemenyi test perfoms pairwise comparisons of the error results for any two architectures. Differences are considered significant if the corresponding average rank differs by at least one critical difference.

\begin{figure}[h]
 \centerline{\includegraphics[width=8cm]{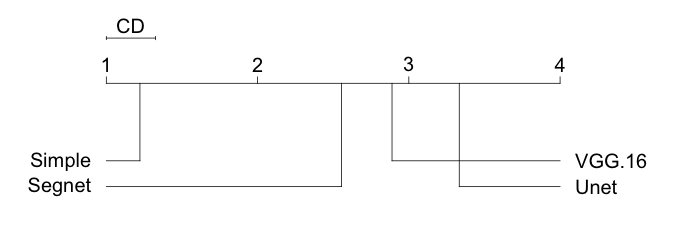}}
  \caption{Critical Differences comparing the 3 convolutional encoder-decoder architectures together the reference simplistic model. $\alpha = 0.05$ }\label{CD}
\end{figure}

Figure \ref{CD} shows a CD diagram representing the results of the Nemenyi test $(\alpha = 0.05)$ using the error values at the nine locations for each convolutional encoder-decoder network. The CD diagrams make a pairwise comparison between methods, connecting the architectures for which no significant statistical differences are found, or in other words, those whose distance is less than the fixed critical difference, shown as reference, at the top of Figure \ref{CD}. Networks ranked with lower values in CD diagrams imply higher error values. These tests have been performed using the \texttt{scmamp} R package, which is publicly available at the Comprehensive R Archive Network (CRAN) \cite{calvo2015}. 

Statistical differences are found between all pairs of networks. As can be seen in the CD diagram, the performance of U-net is significantly better at forecasting total precipitation than the other 3 networks. VGG-16 and Segnet have a significantly lower performance but they are still considerably better than the simple convolutional encoder-decoder described in the first part of the experimental process. These results imply that U-net based architectures provide statistically significant better results when forecasting total precipitation, using geopotential height as input. Considering the results presented in Table \ref{perftable}, U-net requires approximately half the GPU and memory resources than VGG-16 or a quarter than Segnet equivalent networks.

\subsection{Comparison with state-of-the-art machine learning methods}

This section is intended to provide readers with an understanding of the qualitative improvement that deep convolutional architectures offer when compared to other ML methodologies in this context.

First of all, we provide a baseline comparison of precipitation forecast using constant prediction values. We consider two different constant rain fields using zero and the average precipitation over the area of study. These two situations represent the scenarios where we always predict that there is no precipitation or the average precipitation based on the climatology at each grid point. The MAE results of comparing these two scenarios to the ERA-Interim precipitation values over the validation partition are represented in Table \ref{persistence}.

\begin{table}[h]
\caption{Baseline comparison of precipitation forecast using constant values over the whole area.}\label{persistence}
\begin{center}
\begin{tabular}{ll}
$Constant\ value\ [mm]$ & $MAE$\ [mm]\\
 0 (No precipitation) & 0.3417 \\
 0.45 (Mean precipitation) & 0.4845 \\
\end{tabular}
\end{center}
\end{table}

The results in Table \ref{persistence} indicate that forecasting no precipitation provides a substantially better forecast than using the average value, result explained by the fact that in this area most of the grid points do not experience rain at any given time. We refer readers interested in the verification of unlikely event to read the ``Finley effect'' \cite{murphy1996finley}. The mean in this case results in a poor estimate for the precipitation field. The distribution of precipitation has a high variability, precipitations concentrates around well defined clusters and in most of the grid points there is no precipitation. This is why using zero precipitation performs better than the mean value for the MAE metric.

A common technique in computer vision is to train models that learn to predict the value of a pixel using a patch containing the surrounding pixels in the input space \cite{pal2005random,mueller2016water}.
We approach the problem of learning the total precipitation field from the 3 levels of geopotential height determined during the experimental process, but using traditional regression methodologies. We choose three popular regression techniques: linear regression, Least Absolute Shrinkage and Selection Operator (LASSO) and random forest regressor. Due to the high dimensionality of the data, we train the different algorithms using increasingly larger patches (1,3,5,7 and 9) comprising the 3 levels in the input to predict the output's central pixel.

As the size of the patch increases the overlap area between two adjacent patches is larger and the size of the dataset increases, the resulting dataset cannot fit in the memory of high-end machines. To train the different models we randomly sample 100 000 patches of each size.

Table \ref{regress} contain the MAE results of the different regression models for each patch size. This table shows that none of these techniques is capable of learning the relationships between the geopotential and total precipitation fields of a NWP. Also, because these models are trained using a narrow patch or window of the input field, they cannot even match the accuracy of the naive approaches (constant zero and mean precipitation) proposed at the beginning of this section.

\begin{figure*}[h]
 \centerline{\includegraphics[width=16cm]{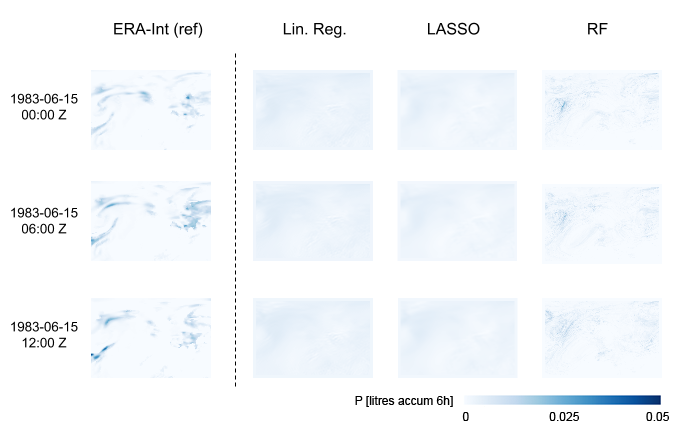}}
  \caption{Output of the three state-of-the-art regression methods (patch size = 7) using the same dates shown in Figure 6.}\label{reg_comp}
\end{figure*}

Figure \ref{reg_comp} shows the output generated by each regression algorithm for the same meteorological situation presented in Figure \ref{comparison}. The models are not able to provide the sharpness necessary to represent the precipitation field. It can be seen that the output generated by random forest provides a light improvement in detecting the position of the precipitation regions, possibly because it is the only non-linear method. However, the capacity of this methodology is not enough to accurately resolve this problem.

\begin{table}[h]
\caption{Comparison of the accuracy level for the different regression models.}\label{regress}
\begin{center}
\begin{tabular}{lccccc}
$method$ & \multicolumn{5}{c}{patch size}\\
 & 1 & 3 & 5 & 7 & 9\\
 Lin.\ Reg. & 0.5281  & 0.5061  & 0.5055  & 0.5054  & 0.5105 \\
 LASSO & 0.5281 & 0.5056 & 0.5049 & 0.5034 & 0.5034\\
 RF & 0.5437 & 0.4924 & 0.4903 & 0.4862 & 0.4851\\
\end{tabular}
\end{center}
\end{table}

\section{Conclusions and future work}

This work demonstrates the suitability of convolutional encoder-decoder networks in learning NWP parameterizations. We performed a series of experiments to demonstrate that these networks can be used to estimate total precipitation, with reasonable accuracy, using only geopotential height field as input. We use the ERA-Interim reanalysis dataset to train different versions of convolutional encoder-decoder networks, using nearly 40 years of geopotential height and total precipitation fields over an extended European region. 

We can conclude, from our experiments, that convolutional encoder-decoder networks are able to extract the relevant information out of the geopotential height field to diagnose total precipitation, as shown in Figures \ref{comparison} and \ref{violin}. Although there are many other physical variables simulated by NWP that contain valuable information to determine the location and intensity of precipitation (ie temperature, humidity), this work's main objective is to demonstrate the capacity of deep neural networks to extract and interpret the synoptic and mesoscale spatial information.

Different architectures of convolutional encoder-decoder networks are statistically compared for a selected group of locations inside the considered region. A modified version of the U-net encoder-decoder network presents the best accuracy, among the considered architectures, at reproducing the total precipitation field. Furthermore we show that deep learning based methods significantly outperform other ML methodologies and traditional methodologies in estimating total precipitation from the geopotential height field. 

NWP precipitation parameterizations consider a combination of several physical fields related to rain formation, such as temperature, humidity or vorticity as inputs. For this work we propose the use of geopotential height at different levels as the only input variable to forecast total precipitation. This is intentional and used to demonstrate the ability of neural networks to find complex non-linear relationships between input and output grid fields. This work also serves as a nod to the community of extremely skilled human weather forecasters that are able to provide an accurate analysis based on the analysis of the 850 and 500 hPa geopotential fields.

In this work we present a series of neural network models that are trained to generate precipitation estimates from the geopotential field. The quality of these models are therefore limited by the quality of the underlying NWP parameterization used to represent the total precipitation field. The same encoder-decoder network could ideally be trained using observed precipitation data instead of parameterized variables. Recent advances in earth observation technologies, such as the Global Precipitation Monitoring (GPM) \cite{hou2014global}, offer high quality global precipitation datasets which could be used in combination with NWP to learn better parameterization models.

Finally, another promising evolution of the methodology presented in this work, is to introduce the temporal dimension to the convolutional encoder-decoder networks using recurrent configurations. Recurrent neural networks \cite{mikolov2010recurrent} have demonstrated remarkable results in the area of time-series analysis and speech recognition and open an interesting new line of research for models that can learn both the spatial and temporal dimensions of NWP data.

\section{acknowledgments}

We would like to thank the National Computational Infrastructure (NCI) at the Australian National University and the University of the Basque Country for their support and advice in carrying out this research work. We are grateful for the support of the Basque Government (IT609-13), the Spanish Ministry of Economy and Competitiveness (TIN2016-78365-R). Jose A. Lozano is also supported by BERC program 2014-2017 (Basque Gov.) and Severo Ochoa Program SEV-2013-0323 (Spanish Ministry of Economy and Competitiveness).

\bibliographystyle{unsrt}  
\bibliography{references} 

\begin{thebibliography}{10}

\bibitem{bauer2015quiet}
Peter Bauer, Alan Thorpe, and Gilbert Brunet.
\newblock The quiet revolution of numerical weather prediction.
\newblock {\em Nature}, 525(7567):47, 2015.

\bibitem{kalnay2003historical}
Eugenia Kalnay.
\newblock Historical overview of numerical weather prediction.
\newblock {\em Handbook of Weather, Climate, and Water: Dynamics, Climate,
  Physical Meteorology, Weather Systems, and Measurements}, pages 95--115,
  2003.

\bibitem{stensrud2009parameterization}
David~J Stensrud.
\newblock {\em Parameterization schemes: keys to understanding numerical
  weather prediction models}.
\newblock Cambridge University Press, 2009.

\bibitem{palmer2005representing}
TN~Palmer, GJ~Shutts, R~Hagedorn, FJ~Doblas-Reyes, Thomas Jung, and
  M~Leutbecher.
\newblock Representing model uncertainty in weather and climate prediction.
\newblock {\em Annu. Rev. Earth Planet. Sci.}, 33:163--193, 2005.

\bibitem{slingo2011uncertainty}
Julia Slingo and Tim Palmer.
\newblock Uncertainty in weather and climate prediction.
\newblock {\em Phil. Trans. R. Soc. A}, 369(1956):4751--4767, 2011.

\bibitem{berner2017stochastic}
Judith et~al Berner.
\newblock Stochastic parameterization: Toward a new view of weather and climate
  models.
\newblock {\em Bulletin of the American Meteorological Society},
  98(3):565--588, 2017.

\bibitem{belochitski2011tree}
Alexei Belochitski, Peter Binev, Ronald DeVore, Michael Fox-Rabinovitz,
  Vladimir Krasnopolsky, and Philipp Lamby.
\newblock Tree approximation of the long wave radiation parameterization in the
  ncar cam global climate model.
\newblock {\em Journal of Computational and Applied Mathematics},
  236(4):447--460, 2011.

\bibitem{o2018using}
Paul~A O'Gorman and John~G Dwyer.
\newblock Using machine learning to parameterize moist convection: Potential
  for modeling of climate, climate change, and extreme events.
\newblock {\em Journal of Advances in Modeling Earth Systems},
  10(10):2548--2563, 2018.

\bibitem{krasnopolsky2013using}
Vladimir~M Krasnopolsky, Michael~S Fox-Rabinovitz, and Alexei~A Belochitski.
\newblock Using ensemble of neural networks to learn stochastic convection
  parameterizations for climate and numerical weather prediction models from
  data simulated by a cloud resolving model.
\newblock {\em Advances in Artificial Neural Systems}, 2013:5, 2013.

\bibitem{brenowitz2018prognostic}
Noah~D Brenowitz and Christopher~S Bretherton.
\newblock Prognostic validation of a neural network unified physics
  parameterization.
\newblock {\em Geophysical Research Letters}, 2018.

\bibitem{lecun2015deep}
Yann LeCun, Yoshua Bengio, and Geoffrey Hinton.
\newblock Deep learning.
\newblock {\em nature}, 521(7553):436, 2015.

\bibitem{dueben2018challenges}
Peter~D Dueben and Peter Bauer.
\newblock Challenges and design choices for global weather and climate models
  based on machine learning.
\newblock {\em Geoscientific Model Development}, 11(10):3999--4009, 2018.

\bibitem{scher2018towards}
S~Scher.
\newblock Towards data-driven weather and climate forecasting: Approximating a
  simple general circulation model with deep learning.
\newblock {\em Geophysical Research Letters}, 2018.

\bibitem{rasp2018deep}
Stephan Rasp, Michael~S Pritchard, and Pierre Gentine.
\newblock Deep learning to represent sub-grid processes in climate models.
\newblock {\em arXiv preprint arXiv:1806.04731}, 2018.

\bibitem{dee2011era}
Dick~P Dee, SM~Uppala, AJ~Simmons, Paul Berrisford, P~Poli, S~Kobayashi,
  U~Andrae, MA~Balmaseda, G~Balsamo, P~Bauer, et~al.
\newblock The era-interim reanalysis: Configuration and performance of the data
  assimilation system.
\newblock {\em Quarterly Journal of the royal meteorological society},
  137(656):553--597, 2011.

\bibitem{lecun2010convolutional}
Yann LeCun, Koray Kavukcuoglu, and Cl{\'e}ment Farabet.
\newblock Convolutional networks and applications in vision.
\newblock In {\em Circuits and Systems (ISCAS), Proceedings of 2010 IEEE
  International Symposium on}, pages 253--256. IEEE, 2010.

\bibitem{bottou2010large}
L{\'e}on Bottou.
\newblock Large-scale machine learning with stochastic gradient descent.
\newblock In {\em Proceedings of COMPSTAT'2010}, pages 177--186. Springer,
  2010.

\bibitem{glorot2010understanding}
Xavier Glorot and Yoshua Bengio.
\newblock Understanding the difficulty of training deep feedforward neural
  networks.
\newblock In {\em Proceedings of the thirteenth international conference on
  artificial intelligence and statistics}, pages 249--256, 2010.

\bibitem{scherer2010evaluation}
Dominik Scherer, Andreas M{\"u}ller, and Sven Behnke.
\newblock Evaluation of pooling operations in convolutional architectures for
  object recognition.
\newblock In {\em Artificial Neural Networks--ICANN 2010}, pages 92--101.
  Springer, 2010.

\bibitem{litjens2017survey}
Geert Litjens, Thijs Kooi, Babak~Ehteshami Bejnordi, Arnaud Arindra~Adiyoso
  Setio, Francesco Ciompi, Mohsen Ghafoorian, Jeroen~AWM van~der Laak, Bram
  Van~Ginneken, and Clara~I S{\'a}nchez.
\newblock A survey on deep learning in medical image analysis.
\newblock {\em Medical image analysis}, 42:60--88, 2017.

\bibitem{baldi2014searching}
Pierre Baldi, Peter Sadowski, and Daniel Whiteson.
\newblock Searching for exotic particles in high-energy physics with deep
  learning.
\newblock {\em Nature communications}, 5:4308, 2014.

\bibitem{dieleman2015rotation}
Sander Dieleman, Kyle~W Willett, and Joni Dambre.
\newblock Rotation-invariant convolutional neural networks for galaxy
  morphology prediction.
\newblock {\em Monthly notices of the royal astronomical society},
  450(2):1441--1459, 2015.

\bibitem{hinton2006reducing}
Geoffrey~E Hinton and Ruslan~R Salakhutdinov.
\newblock Reducing the dimensionality of data with neural networks.
\newblock {\em science}, 313(5786):504--507, 2006.

\bibitem{masci2011stacked}
Jonathan Masci, Ueli Meier, Dan Cire{\c{s}}an, and J{\"u}rgen Schmidhuber.
\newblock Stacked convolutional auto-encoders for hierarchical feature
  extraction.
\newblock In {\em International Conference on Artificial Neural Networks},
  pages 52--59. Springer, 2011.

\bibitem{krizhevsky2012imagenet}
Alex Krizhevsky, Ilya Sutskever, and Geoffrey~E Hinton.
\newblock Imagenet classification with deep convolutional neural networks.
\newblock In {\em Advances in neural information processing systems}, pages
  1097--1105, 2012.

\bibitem{long2015fully}
Jonathan Long, Evan Shelhamer, and Trevor Darrell.
\newblock Fully convolutional networks for semantic segmentation.
\newblock In {\em Proceedings of the IEEE conference on computer vision and
  pattern recognition}, pages 3431--3440, 2015.

\bibitem{chen2018deeplab}
Liang-Chieh Chen, George Papandreou, Iasonas Kokkinos, Kevin Murphy, and Alan~L
  Yuille.
\newblock Deeplab: Semantic image segmentation with deep convolutional nets,
  atrous convolution, and fully connected crfs.
\newblock {\em IEEE transactions on pattern analysis and machine intelligence},
  40(4):834--848, 2018.

\bibitem{isola2017image}
Phillip Isola, Jun-Yan Zhu, Tinghui Zhou, and Alexei~A Efros.
\newblock Image-to-image translation with conditional adversarial networks.
\newblock {\em arXiv preprint}, 2017.

\bibitem{badrinarayanan2017segnet}
Vijay Badrinarayanan, Alex Kendall, and Roberto Cipolla.
\newblock Segnet: A deep convolutional encoder-decoder architecture for image
  segmentation.
\newblock {\em IEEE Transactions on Pattern Analysis and Machine Intelligence},
  39(12):2481--2495, 2017.

\bibitem{ronneberger2015u}
Olaf Ronneberger, Philipp Fischer, and Thomas Brox.
\newblock U-net: Convolutional networks for biomedical image segmentation.
\newblock In {\em International Conference on Medical image computing and
  computer-assisted intervention}, pages 234--241. Springer, 2015.

\bibitem{1321008426928}
P.~Berrisford, D.P. Dee, P.~Poli, R.~Brugge, K.~Fielding, M.~Fuentes, P.W. K{\r
  a}llberg, S.~Kobayashi, S.~Uppala, and A.~Simmons.
\newblock The era-interim archive version 2.0.
\newblock Shinfield Park, Reading, November 2011.

\bibitem{saeys2007review}
Yvan Saeys, I{\~n}aki Inza, and Pedro Larra{\~n}aga.
\newblock A review of feature selection techniques in bioinformatics.
\newblock {\em bioinformatics}, 23(19):2507--2517, 2007.

\bibitem{reunanen2003overfitting}
Juha Reunanen.
\newblock Overfitting in making comparisons between variable selection methods.
\newblock {\em Journal of Machine Learning Research}, 3(Mar):1371--1382, 2003.

\bibitem{chollet2017keras}
Fran{\c{c}}ois Chollet et~al.
\newblock Keras (2015), 2017.

\bibitem{abadi2016tensorflow}
Mart{\'\i}n Abadi, Paul Barham, Jianmin Chen, Zhifeng Chen, Andy Davis, Jeffrey
  Dean, Matthieu Devin, Sanjay Ghemawat, Geoffrey Irving, Michael Isard, et~al.
\newblock Tensorflow: A system for large-scale machine learning.
\newblock In {\em OSDI}, volume~16, pages 265--283, 2016.

\bibitem{sanders1995case}
Frederick Sanders and Charles~A Doswell~III.
\newblock A case for detailed surface analysis.
\newblock {\em Bulletin of the American Meteorological Society},
  76(4):505--522, 1995.

\bibitem{sanders1999proposed}
Frederick Sanders.
\newblock A proposed method of surface map analysis.
\newblock {\em Monthly weather review}, 127(6):945--955, 1999.

\bibitem{tensorflow2015-whitepaper}
Mart\'{\i}n Abadi, Ashish Agarwal, Paul Barham, Eugene Brevdo, Zhifeng Chen,
  Craig Citro, Greg~S. Corrado, Andy Davis, Jeffrey Dean, Matthieu Devin,
  Sanjay Ghemawat, Ian Goodfellow, Andrew Harp, Geoffrey Irving, Michael Isard,
  Yangqing Jia, Rafal Jozefowicz, Lukasz Kaiser, Manjunath Kudlur, Josh
  Levenberg, Dandelion Man\'{e}, Rajat Monga, Sherry Moore, Derek Murray, Chris
  Olah, Mike Schuster, Jonathon Shlens, Benoit Steiner, Ilya Sutskever, Kunal
  Talwar, Paul Tucker, Vincent Vanhoucke, Vijay Vasudevan, Fernanda Vi\'{e}gas,
  Oriol Vinyals, Pete Warden, Martin Wattenberg, Martin Wicke, Yuan Yu, and
  Xiaoqiang Zheng.
\newblock {TensorFlow}: Large-scale machine learning on heterogeneous systems,
  2015.
\newblock Software available from tensorflow.org.

\bibitem{hintze1998violin}
Jerry~L Hintze and Ray~D Nelson.
\newblock Violin plots: a box plot-density trace synergism.
\newblock {\em The American Statistician}, 52(2):181--184, 1998.

\bibitem{demsar2006}
J.~Demsar.
\newblock Statistical comparisons of classifiers over multiple data sets.
\newblock {\em Journal of Machine Learning Research}, 7:1--30, 2006.

\bibitem{pohlert2014pairwise}
Thorsten Pohlert.
\newblock The pairwise multiple comparison of mean ranks package (pmcmr).
\newblock {\em R package}, pages 2004--2006, 2014.

\bibitem{calvo2015}
B.~Calvo and G.~Santafe.
\newblock scmamp: Statistical comparison of multiple algorithms in multiple
  problems.
\newblock {\em The R Journal}, 2016.

\bibitem{murphy1996finley}
Allan~H Murphy.
\newblock The finley affair: A signal event in the history of forecast
  verification.
\newblock {\em Weather and Forecasting}, 11(1):3--20, 1996.

\bibitem{pal2005random}
Mahesh Pal.
\newblock Random forest classifier for remote sensing classification.
\newblock {\em International Journal of Remote Sensing}, 26(1):217--222, 2005.

\bibitem{mueller2016water}
Norman Mueller, Adam Lewis, Dale Roberts, Steven Ring, R~Melrose, J~Sixsmith,
  Leo Lymburner, A~McIntyre, P~Tan, S~Curnow, et~al.
\newblock Water observations from space: Mapping surface water from 25 years of
  landsat imagery across australia.
\newblock {\em Remote Sensing of Environment}, 174:341--352, 2016.

\bibitem{hou2014global}
Arthur~Y Hou, Ramesh~K Kakar, Steven Neeck, Ardeshir~A Azarbarzin, Christian~D
  Kummerow, Masahiro Kojima, Riko Oki, Kenji Nakamura, and Toshio Iguchi.
\newblock The global precipitation measurement mission.
\newblock {\em Bulletin of the American Meteorological Society},
  95(5):701--722, 2014.

\bibitem{mikolov2010recurrent}
Tom{\'a}{\v{s}} Mikolov, Martin Karafi{\'a}t, Luk{\'a}{\v{s}} Burget, Jan
  {\v{C}}ernock{\`y}, and Sanjeev Khudanpur.
\newblock Recurrent neural network based language model.
\newblock In {\em Eleventh Annual Conference of the International Speech
  Communication Association}, 2010.

\end{thebibliography}
\end{document}